\documentclass[ras]{agutex}

\usepackage[dvipdf]{graphicx}
\authorrunninghead{LANE ET AL.}

\titlerunninghead{VLSS REDUX}

\authoraddr{W. M. Lane, Naval Research Laboratory, Remote Sensing
  Division, Code 7213, 4555 Overlook Ave. SW, Washington, DC 23185,
  USA. (wendy.peters@nrl.navy.mil)}

\authoraddr{W. D. Cotton, National Radio Astronomy Observatory, 520 Edgemont Road, Charlottesville, VA,  22903, USA.}

\authoraddr{J. F. Helmboldt, Naval Research Laboratory, Remote Sensing
  Division, Code 7213, 4555 Overlook Ave. SW, Washington, DC 23185,
  USA.}

\authoraddr{N. E. Kassim, Naval Research Laboratory, Remote Sensing
  Division, Code 7213, 4555 Overlook Ave. SW, Washington, DC 23185,
  USA.}

\begin{document}

%% ------------------------------------------------------------------------ %%
%
%  TITLE
%
%% ------------------------------------------------------------------------ %%

\title {VLSS Redux: Software Improvements applied to the Very Large Array
  Low-frequency Sky Survey}

%
% e.g., \title{Terrestrial ring current:
% Origin, formation, and decay $\alpha\beta\Gamma\Delta$}
%

%% ------------------------------------------------------------------------ %%
%
%  AUTHORS AND AFFILIATIONS
%
%% ------------------------------------------------------------------------ %%

%Use \author{\altaffilmark{}} and \altaffiltext{}

% \altaffilmark will produce footnote;
% matching \altaffiltext will appear at bottom of page.

\authors{W. M. Lane,\altaffilmark{1} W. D. Cotton,\altaffilmark{2}
  J. F. Helmboldt,\altaffilmark{1} N. E. Kassim,\altaffilmark{1}}

\altaffiltext{1}{Naval Research Laboratory, Code 7213, 4555 Overlook Ave. SW, Washington, DC, 20375, USA.}

\altaffiltext{2}{National Radio Astronomy Observatory, 520 Edgemont Road, Charlottesville, VA,  22903, USA.}

%% ------------------------------------------------------------------------ %%
%
%  ABSTRACT
%
%% ------------------------------------------------------------------------ %%

% >> Do NOT include any \begin...\end commands within
% >> the body of the abstract.

\begin{abstract}

We present details of improvements to data processing and analysis
which were recently used for a re-reduction of the Very Large Array
(VLA) Low-frequency Sky Survey (VLSS) data.  Algorithms described are
implemented in the data-reduction package Obit, and include
smart-windowing to reduce clean bias, improved automatic radio
frequency interference removal, improved bright-source peeling, and
higher-order Zernike fits to model the ionospheric phase
contributions.  An additional, but less technical improvement was
using the original VLSS catalog as a same-frequency/same-resolution
reference for calculating ionospheric corrections, allowing more
accuracy and a higher percentage of data for which solutions are
found.  We also discuss new algorithms for extracting a source catalog
and analyzing ionospheric fluctuations present in the data.  The
improved reduction techniques led to substantial improvements
including images of six previously unpublished fields (1$\%$ of the
survey area) and reducing the clean bias by 50$\%$.  The largest angular
size imaged has been roughly doubled, and the number of cataloged
sources is increased by 35$\%$ to 95,000.

\end{abstract}

%% ------------------------------------------------------------------------ %%
%
%  BEGIN ARTICLE
%
%% ------------------------------------------------------------------------ %%

% The body of the article must start with a \begin{article} command
%
% \end{article} must follow the references section, before the figures
%  and tables.

\begin{article}

%% ------------------------------------------------------------------------ %%
%
%  TEXT
%
%% ------------------------------------------------------------------------ %%

\section{Introduction}

The Very Large Array (VLA) Low-frequency Sky Survey (VLSS), released
in ~\citet{VLSS1} covers 95$\%$ of the $3\pi$ sr of sky area above
-30\deg\ declination at a frequency of 74 MHz, a resolution of
approximately $80''$, and an RMS sensitivity of $\approx 0.1$ Jy/bm.
The main survey products consist of a publicly available catalog and a
set of maps.  The survey was intended to serve as a low-frequency
counterpart to the National Radio Astronomy Observatory (NRAO)-VLA Sky
Survey (NVSS) at 1400 MHz ~\citep{NVSS}, allowing spectral information
to be compiled for statistical samples of sources.  It also provides a
low-frequency sky model.

The original data reduction was hampered by limited software.  In the
past few years, several major improvements to the processing software
along with the availability of faster computers which could process
the data in a fraction of the time originally needed, made it
attractive to re-reduce the survey data.  The goal of the re-reduction
was to increase the sensitivity and uniformity of the survey and maps.
The pipeline data processing developed could also be leveraged as a
basis for future low frequency data reduction.

In addition to software limitations, one of the most significant
limitations to the original VLSS data reduction was the lack of a sky
model at a comparable frequency.  In order to calculate the Zernike
polynomials for the phase screen to correct the variable ionosphere
across the field of view, a sky model was extrapolated from the 1400
MHz NVSS using an assumed standard spectral index of $\alpha = -0.7$.
This extrapolated sky model was adequate but led to many false
results, where sidelobes of other sources were picked up instead of
the real source, which might be much fainter than anticipated; at the
same time many steeper spectrum sources which could have been used to
improve the solution fits were not included.  When considering a
re-reduction we quickly realized that we could use the original VLSS
catalog itself for a sky model.  With no need to estimate source flux,
we could focus on true sources in the Zernike fitting, leading to
cleaner fits and better solutions.

We have reprocessed all of the VLSS data from the archive to make new
maps and a new catalog. The details of the VLSS Redux (VLSSr) maps and
catalogs, will be described in a separate paper.  Here we discuss
improvements made to the basic data reduction and analysis.  In
Section 3 of this paper we discuss the data processing including:
``smart-windowing'' to reduce clean-bias, automated radio frequency
interference modeling software, a revised peeling method, and improved
ionospheric phase corrections.  In Section 4 we present a new ``false
detection rate'' limited cataloging method, and improved ionospheric
fluctuation calculations.  All of the processing described is
implemented in the data reduction package Obit \citep{OBIT}, except
the ionospheric fluctuations analysis which makes use of additional
independent software.

\section{The Reprocessing}

Here we briefly describe the steps of the reprocessing.

The initial calibration of the data was done in the Astronomical Image
Processing Software (AIPS) and remained as described in
~\citet{VLSS1}.  The only change was to eliminate the data-editing
steps intended to remove radio frequency interference (RFI); aside
from a global clip of very high amplitude data points no editing was
done in the initial calibration.

The imaging was a three-step process for each of the 523 pointing
centers in the survey.  The data were corrected for ionospheric
distortions and imaged, and a residual data set with no astronomical
signal was produced.  Using these residuals the RFI was modeled and
that model was removed from the original data set.  Corrections for
the ionospheric distortions were re-calculated using the RFI-corrected
data set and a final field map made.  Offset information for the
calibrators was kept for use in ionospheric fluctuation analysis.

The pointing center maps were weighted by 1/RMS and combined to create
mosaic image squares, in which the overlap of the pointings produces a
more uniform sensitivity.

The squares were cataloged by fitting Gaussians to peaks above a given
detection level.  We cataloged the survey both using the traditional,
local 5$\sigma$ catalog limit, and also using a new method based on
predicted false detection rate.

\section{Data Reprocessing Improvements}
\subsection{Field--based Ionospheric Correction}
\subsubsection{Background}

An electromagnetic wavefront passing through the ionosphere will
encounter a space and time variable refractive index, which is mainly
due to the variable free electron density. A wedge in the integrated
electron density (total electron content, or TEC) along the wave's
trajectory will cause a linear phase gradient across an array
observing through it, resulting in an apparent position shift for any
small source in the field of view.  The apparent source position
shifts are proportional to the TEC gradient in a given direction and
thus may vary across the field of view of the array elements. Higher
order phase structures across the array cause a more serious
distortion of the wavefront, producing source defocusing, and in
extreme cases scintillations \citep{Lonsdale}.

In the regime of linear phase gradients, the "field-based" ionospheric
correction method is applicable; it has been described in detail in
~\citet{Cotton2004SPIE} and ~\citet{Cotton2005WCE}.  The technique is
to make a series of snapshot measurements around the locations of
known strong sources (calibrators) in the field, deconvolve the
images, and estimate the apparent offsets of each.  The time sequence
of the derived set of source position offsets allows the fitting of a
time variable geometric distortion of the sky as seen by the array.
Low order Zernike polynomials, which are orthogonal on a circle, are
used to model the distortion field.  The field is modeled as a phase
screen and each position offset measurement gives a 2-D gradient in
this screen at the ionospheric puncture point of the line of sight to
the calibrator.

At low frequencies with 2-D arrays, some provision must be made for
array non-coplanarity ~\citep{Cornwell1992}.  One solution to this
problem is the ``Fly's eye'' approach where the sky is tiled with many
small facets, each tangent to the celestial sphere at its center.  In
practice, the size of the tile needed is smaller than the isoplanatic
patch size (the characteristic scale over which the rms phase
difference between two lines of sight is approximately 1 rad,
equivalent to a linear size of a few tens of km at 74 MHz
\citep{Cotton2004SPIE}) and/or the resolution at which the phase
screen can be determined, so a sufficient approximation to
de-distorting the sky is to correct each facet for the geometric
offset at its center.  This is done by calculating the antenna-based
phase corrections at the center of the facet and applying these
corrections prior to deriving the dirty image (or residual) of that
facet.

An initial Zernike fit is made to each time segment, in which the
source offsets are allowed to be arbitrarily large (the limit is set
by the user; for the VLSSr 10' was used).  This initial distortion
model is used to refine the expected source positions.  The calibrator
offsets are then recomputed using the adjusted model positions.
Calibrators are then required to be found within a small radius of the
expected position (a sum of 10 pixels plus $10\%$ of the tip-tilt term
of the Zernike fit plus the rms of the initial fit).  Calibrators with
offsets greater than this radius are excluded from further fitting.

An adjustment to the model is made to compensate for possible real
differences between the model positions and the data positions (likely
for extended sources if the input catalog is not at the same
frequency/resolution as the data).  The model positions of sources
which are flagged in the input catalog as being either resolved or
having close neighboring sources are adjusted if the average offset
residual for the calibrator over all time intervals exceeds half of
the residual RMS for all sources.  Calibrators which are both isolated
and unresolved in the model are not adjusted.

The fits to the individual time segments are then recomputed and the
average residual offset is compared to a target RMS residual, which is
essentially the residual "seeing" size that is allowed; in practice we
find that a quarter to a third of the synthesized beam is a good
choice.  If the average RMS of the residuals is greater than the
target, the most discrepant remaining calibrator offset which has at
least 1.5 times the average variance is rejected and the Zernike fit
is recomputed.  This is repeated until one of the following conditions
is met: 1) the RMS residual is acceptable, 2) there is no calibrator which
contributes more than 1.5 times the average variance, or 3) there are too
few measurements for a fit.  In the latter two cases, this time
segment is flagged and excluded from further imaging.

The ionosphere can be extremely variable and at times the data cannot
be adequately corrected for its effect with the Zernike fits; these
times should be excluded from imaging.  They are usually indicated by
defocusing; in extreme cases, none of the calibrators can be detected
so no field-based calibration is possible.  In less extreme cases, the
sources are still detectable so they pass the field-based calibration
step.  For these data, defocusing can be identified from the peak image
values of the calibrators.  For each calibrator, the average image
peak is determined.  If in any time segment, the average ratio of the
calibrator peak to its average drops below $50\%$, the time segment is
rejected.  $50\%$ was found to be a good compromise in general between
removing too much data and keeping poor data when we first started
reducing 74 MHz VLA data, but the parameter can be changed in the
software if desired.  The remaining time sequence of fitted Zernike
polynomials is applied in the imaging and deconvolution as was
described in ~\citet{VLSS1}.

\subsubsection{Improvements for the VLSSr \label{sec:ioncalr}}
The field-based calibration used in the processing of the VLSSr
differs from the original processing in a number of respects.
Principal among these are using the source catalog from the VLSS as the
calibrator list and using a higher order Zernike model.

The original VLSS field-based calibration used the NVSS as a sky
model.  Because this is at a very different frequency and resolution
from the data, it was necessary to predict 74 MHz flux values for the
sources using an average spectral index of $\alpha = -0.7$.  However
it was not possible to tell which of the potential calibrator sources
would actually be present in the data at that flux.  For any
calibrator, the true source might not be detectable, and false
detections of sidelobes instead of true sources contaminated the
calibrator sample used in the Zernike calculations.  The high
probability of a false detection made it necessary to limit how far
from the nominal source position we searched, and thus times with
larger ionospheric disturbances were lost.  This problem was almost
completely eliminated in the VLSSr by using the original VLSS source
catalog ~\citep{VLSS1} for the sky model.  By using a sky-model at the
same frequency and resolution, we are certain that every calibrator
source exists at the expected flux in the data, and can therefore
include more sources and search for them over a wider shift area
without compromising the solutions.  There were a few areas on the sky
where the original VLSS was incomplete or insufficient for good
ionospheric calibration (roughly 2$\%$ of the fields); in those cases
the NVSS was used.

Because there are a greatly expanded set of reliable calibrators in
the sky model, further refinements to the calibrator selection can be
made to improve the quality of the Zernike model fits.  First, any
calibrator measurement in which the integrated value is less than a
third or more than three times the peak is rejected to remove heavily
resolved sources and sidelobes.  The next level of filtering is to
make a preliminary fit of the Zernike model and restrict the
calibrator set to those sources with offsets which do not grossly
differ from that model.  The Zernike model is then re-fit to the
offsets of the final selection of calibrator sources, with all fits
weighted by the calibrator peak flux density.

The improved initial sky model and subsequent selection criteria allow
the inclusion of more measurements of the ionospheric gradient over a
wider range of spatial scales (see Section ~\ref{sec:ion}) than in the
original reduction.

In the original VLSS we included sources stronger than a predicted
flux of 3 Jy (extrapolated from the NVSS), and were, for most fields,
forced to model the ionosphere at 2 minute intervals to improve the
dynamic rang in the offset measurements.  The number of sources we
could find to model each time interval were few enough that the
Zernike solutions were limited to 2nd-order polynomials, and a large
fraction of the data was lost when a good solution could not be found
during a given time interval.

By contrast, for the VLSSr we were able to reliably use sources with
measured total flux of 2.5 Jy or greater and solve for the Zernike
polynomials at 1 minute intervals.  We ran both 2nd and 3rd order
Zernike solutions for all fields.  For roughly 70$\%$ of the fields,
the 3rd order solution produced a ``better'' map based on the criteria
of a higher dynamic range and greater maximum peak flux; we also made
sure the two maps had comparable total flux in the field (no sources
were lost or power scattered around by one of the two calibration
methods).  Visual inspection was made of any field where these
criteria did not clearly indicate a better map and/or where the total
flux values were not comparable.

\begin{figure}[t]
\noindent\includegraphics[width=18pc]{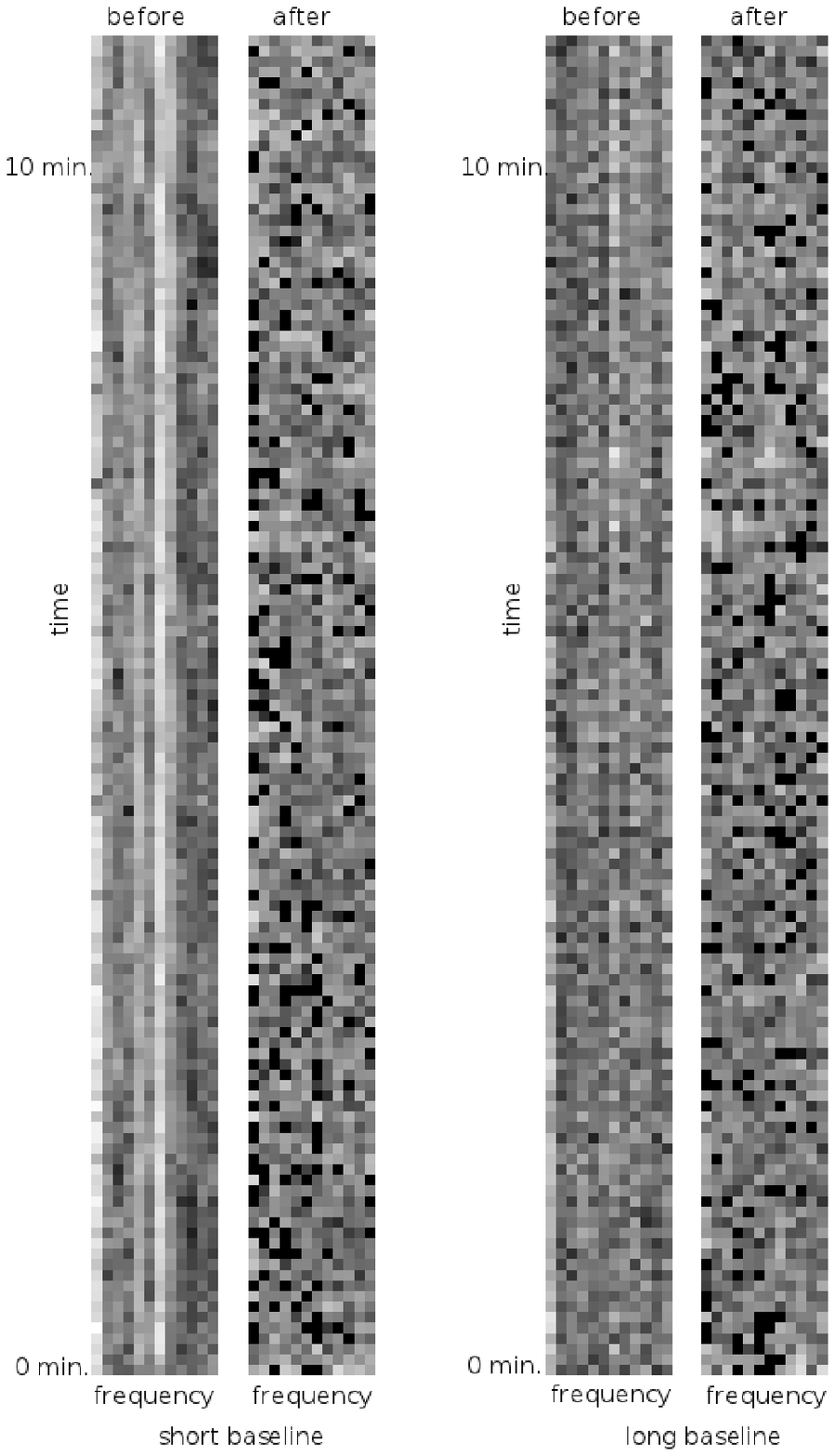}
\caption{\center A short fragment of data on a single baseline is show before
  and after applying the RFI removal as described in the text.  The
  baseline on the left is a short baseline (E32-E20) while the data on
  the right is for a longer baseline (E32-W32). The greyscale is
  auto-scaled in arbitrary units, with white indicating larger flux
  values. \label{fig:rfi} }
\end{figure}

\subsection{RFI Excision}

Radio Frequency Interference (RFI) is a persistent problem at lower radio
frequencies and can seriously corrupt images.  Many of the
interfering signals are broadband and/or slowly varying in time making
them more difficult to detect than impulsive or narrow band signals.
The RFI mitigation strategy used for the VLSSr is a combination of the
traditional "flagging" of the most seriously affected data coupled
with an RFI estimation and subtraction technique similar to the one
described by ~\citet{Athreya2009}.

Initial editing of the data removes any visibility measurements with
extremely large amplitudes.  This allows the data to be imaged and an
initial model of the sky to be subtracted from the data.  The residual
data should be dominated by the RFI and can be used to estimate the
effect of RFI on the data; by working on residuals we minimize the
chance of removing any real celestial signals during the process.  

Stationary terrestrial-based interfering signals should have a
constant phase as seen by the array, whereas celestial signals will
have a phase which is constantly varying due to the changing geometry
caused by the rotation of the earth.  This earth rotation induced
phase variation is removed in the correlation by a process known as
"phase tracking".  The phase tracking will cause a celestial point
source at the phase tracking position to have a constant phase whereas
any terrestrial RFI will have a variable phase.  This process can be
reversed for the residual data, counter-rotating the data by the
inverse of the phase tracking.  This will cause the terrestrial RFI to
have a constant phase and any remaining celestial signals to rotate.
Time averaging of this counter-rotated data will further smear out any
residual celestial emission but leave constant RFI unaffected.

The averaged counter-rotated residual data can then be filtered to
form a time variable model of the RFI.  RFI will not be present at all
times and baselines so only values above a minimum threshold are
accepted in the RFI model.  The resulting RFI model then has the phase
tracking re-applied.  It is interpolated to the data sampling times
for each baseline, frequency, and polarization and subtracted from the
data.  For very short baselines, the difference in the phase rotation
of celestial and terrestrial sources may not be sufficiently large to
separate them; to compensate, these data are removed completely if
they exceed the minimum RFI threshold.  The model is subtracted from
the original data to produce a data set containing the celestial
signals but with the estimate of the RFI removed.

The RFI modeled by the process described above is not always
sufficiently constant in time that it can be completely removed by
this technique.  To compensate for this, the RFI model is also
subtracted from the residual data and the times, baselines,
frequencies and polarizations of any values with amplitudes above
nominal values in I and V are excluded from further processing.  Any
baseline, channel or IF which has more than 25$\%$ of its data
excluded by the stokes V test is removed completely.  Each baseline is
further filtered in the frequency domain by removing frequency
channels with an RMS that differs by a chosen amount from the median
level during a given time interval. The edited, RFI subtracted data is
then ready to be re-imaged.

The RFI estimation process is implemented in Obit task LowFRFI and is
described in more detail in ~\citet{CottonRFIMemo}; the subsequent
data clipping is implemented in Obit task AutoFlag.  

For the VLSSr, we found the following parameters gave good results in
our early tests.  Initial data editing was done in AIPS to remove all
visibilities with amplitudes greater than two times the zero-spacing
flux, as estimated by fitting for flux vs. UV-distance and
extrapolating back to a distance of zero.  For the RFI modeling, the
data were averaged for 8 minutes and the minimum RFI amplitude
threshold was 0.5 Jy.  For the subsequent editing step, data with
stokes I flux $> 400$ Jy or stokes V flux $> 300$ Jy were removed.
For each 10 second sample, frequency channels with an RMS which
differed from the median of all channels by more than $6\sigma$ were
removed.

In the original VLSS we completely removed channels which were part of
the 100kHz interference ``comb'' generated by the VLA itself
~\citep{VLA74}; however this frequently removed good data as the comb
did not appear at equal strengths on all baselines.  For the VLSSr we
let the RFI modeling algorithm remove the comb. Figure 1 shows VLSSr
data on two sample baselines before and after applying the RFI removal
steps described here.  For the short baseline, although the RFI
dominates much of the frequency band, most of the data were able to be
retained.  Although initially there is far more RFI structure on the
short baseline, the RFI-subtracted and flagged data look very similar
on both the long and short baselines, without visible interference,
and without the necessity of excising large portions of the short
baseline.

The original VLSS excluded all baselines shorter than 200$\lambda$
from the processing to reduce RFI.  This limited the theoretical
largest angular scale of the survey to $18'$.  By using the RFI
modeling and removal techniques described here we were able to include
all baselines present in the data.  This doubles the theoretical
largest angular scale in the survey to $\approx 36'$.  Because the
data do not have complete UV-coverage of any field the actual largest
angular scale is lower in both reductions.  The increase in extended
source sensitivity has a dramatic impact on the appearance of large
sources, such as Galactic supernova remnants.  Combined with the lower
noise values, we also see some new large-scale features in
extragalactic sources, such as the radio tails of the galaxies in the
center of the cluster A194.  Images of two large objects are shown in
Figure 2 to illustrate the improvements achieved by the new
processing.

\begin{figure*}[t]
\noindent\includegraphics[width=39pc]{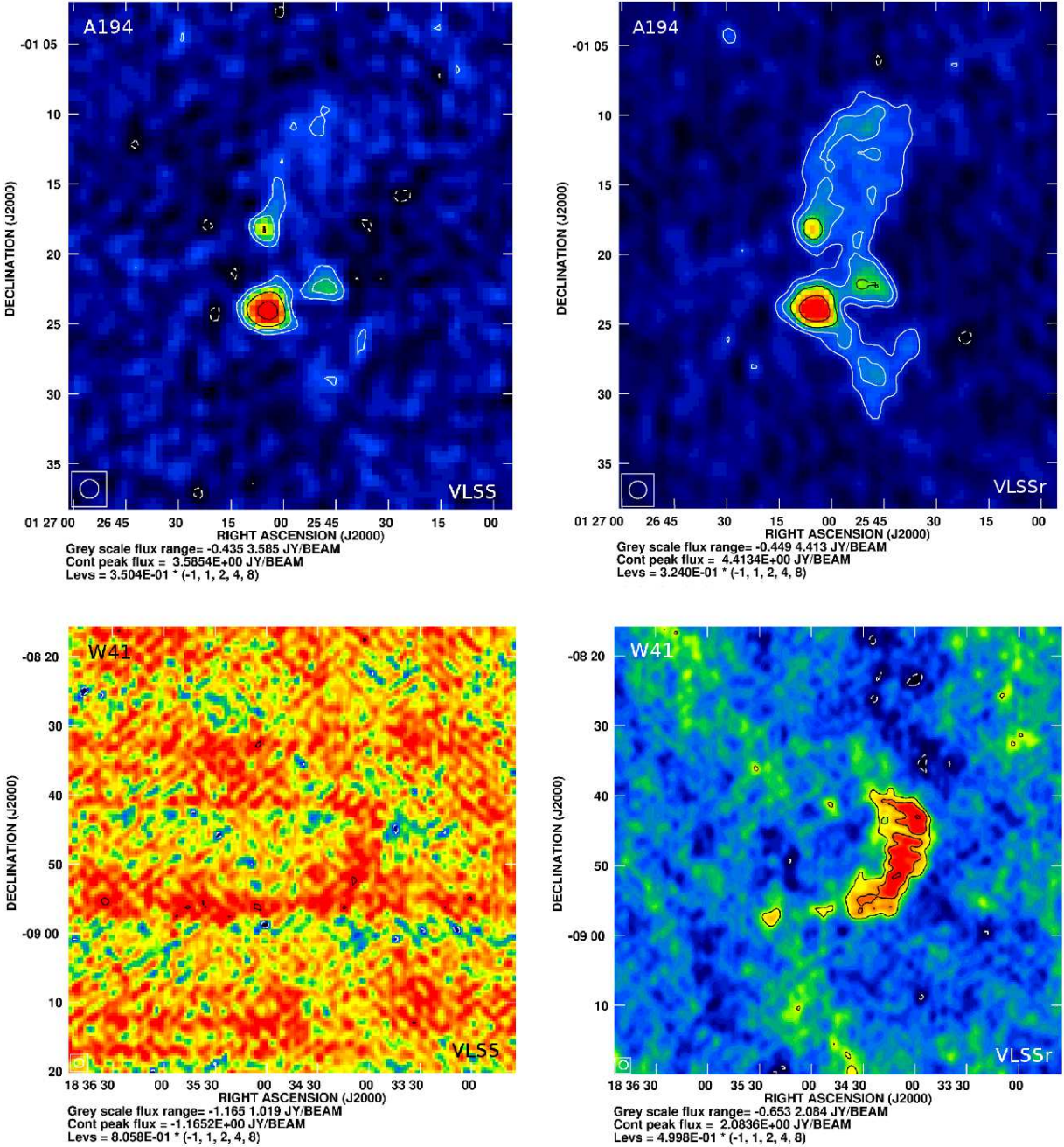}
\caption{Comparison of the VLSS ({\it left}) and VLSSr ({\it right})
  for two extended objects.  Abell 194 ({\it top}) is two luminous and
  distorted radio galaxies at the center of a low-redshift galaxy
  cluster. The improvement is due to a combination of increased
  extended source sensitivity and lower noise.  W41 ({\it bottom}) is
  a giant shell-type supernova remnant in the Milky Way Galaxy.  The
  improvement is due to the improved RFI suppression and the increased
  extended source sensitivity.  Because the VLSS and VLSSr images have
  slightly different restoring beams and the pixel values are given in
  Jy/beam, the objects are plotted at equivalent, rather than
  identical, flux scale ranges based on the minimum and maximum pixel
  value in each image.  Images are contoured at multiples of
  $3\sigma$. \label{fig:pretty}}
\end{figure*}

\subsection{Peeling}
Peeling is a term used to mean the calibration, imaging and removal of
one source from a data set, with the goal of better imaging the
remaining sources.  By removing the source completely, peeling
effectively also removes the sidelobes of the source, which can be
important for imaging weak sources in the presence of a very strong
source.  When many sources are peeled in sequence, it can be used to
build up a large scale image where each source has been individually
calibrated to remove ionospheric phase terms that are variable across
the field of view ~\citep{IntemaSPAM}.

Because we had the ability to use field-based calibration to model and
correct position-variable ionospheric phase terms, we used peeling
only to mitigate the effects of sidelobes from bright sources in the
VLSSr.  Direction dependent calibration focuses the field of view over
which the solutions are valid.  If the distribution of calibrators
used in the solutions is not optimal so that the phase screen is not
calculated over the entire imaging area, or if the ionospheric phase
screen is more complicated than can be described by the polynomials,
sources may not be ideally focused, and may retain sidelobes.  This is
particularly true for sources which are not well-centered on a pixel
and for sources near the edges of the image.  For most sources these
are below the RMS noise in the image and can be ignored; however for
bright sources they may leave imaging artifiacts that we wish to
remove by peeling.

The peeling algorithm is included in the Obit task ``IonImage''.  A
Zernike-based ionospheric phase model is derived using all the data
and an initial image is made.  If any sources in the image have a peak
greater than the chosen limit, they are peeled.  All other sources are
subtracted from a temporary copy of the dataset and a small image is
centered at the bright source position so the source is centered on a
pixel, The data undergo several loops of phase self-calibration and
re-imaging. For the last loop the data are amplitude and phase
self-calibrated before producing a final image and clean-component
model.  The model is then distorted by the inverse of the calculated
self-calibration solutions and subtracted from the original
uncalibrated UV-data.  The subtracted data are re-imaged and the final
peeled source model is reinserted into the map at the end.

The key improvement to this algorithm is distorting the model with the
self-calibration solutions rather than the entire data set.
Calibrating and then uncalibrating the entire data set to peel a
source introduced small errors each time it was done and greatly
limited the effectiveness of peeling in the original VLSS survey. As a
result, only a few of the very brightest sources on the sky could be
peeled.  For the VLSSr, we found image improvements with peeling to
much lower levels based on inspection of a subset of fields.  All
sources with a peak flux $>25$Jy were peeled to reduce sidelobe
levels.

\subsection{Smart-Window Cleaning}
Images deconvolved using the CLEAN algorithm
~\citep{hogbom74,schwarz78} are known to suffer from a ``clean bias''
which systematically reduces the flux of sources in the field (see
\citet{NVSS} for a description).  This occurs because, as cleaning
proceeds to deeper levels, the probability increases that a sidelobe
of a source or a noise fluctuation (or a combination of both) can
produce a peak higher than any remaining flux in the image.  Cleaning
this false source results in flux from its modeled sidelobes being
subtracted from the true sources in the field.  Therefore, the clean
bias results in the flux densities of sources being systematically
reduced.  The magnitude of the bias is independent of the flux density
of sources, but scales with map noise.

One way to reduce the amount of clean bias introduced is to clean only
in small areas focused on known real sources.  However it can be
tedious to set up hundreds of windows around sources in a
well-populated low-frequency field; and even harder to know {\it a
  priori} how large each window needs to be to include the source but
not the surrounding noise.

The Obit task ``Ion Image'' includes a ``smart-windowing'' system which
attempts to automatically determine where to clean
~\citep{autoWindow}. A new box is added to the CLEAN window if the
peak residual in a given facet is inside the CLEANable region, but
outside the current window, and the peak exceeds five times the
residual RMS. The one dimensional structure function of the residual
pixel values is then evaluated to determine the size of a round box
centered on the peak. The box is given a radius at which the square
root of the structure function drops to the greater of 10\% of the
peak or three times the residual RMS.

While this smart-windowing cannot completely remove clean bias, it
does greatly diminish it.  Clean bias scales with the local noise, and
can therefore be expressed as a multiple of the local noise.  As
discussed in ~\citet{OBIT}, tests using a CLEAN process which is
constrained to not clean weak sources deeply show that the windowing
can reduce the clean bias to as little as $0.2\sigma$.  For the VLSSr
using the windowing reduced the clean bias for point sources in our
maps by over $50\%$ from $1.39\sigma$ in the original published VLSS
to $0.66\sigma$ in the VLSSr. 

\begin{figure}[t]
\noindent\includegraphics[width=18pc]{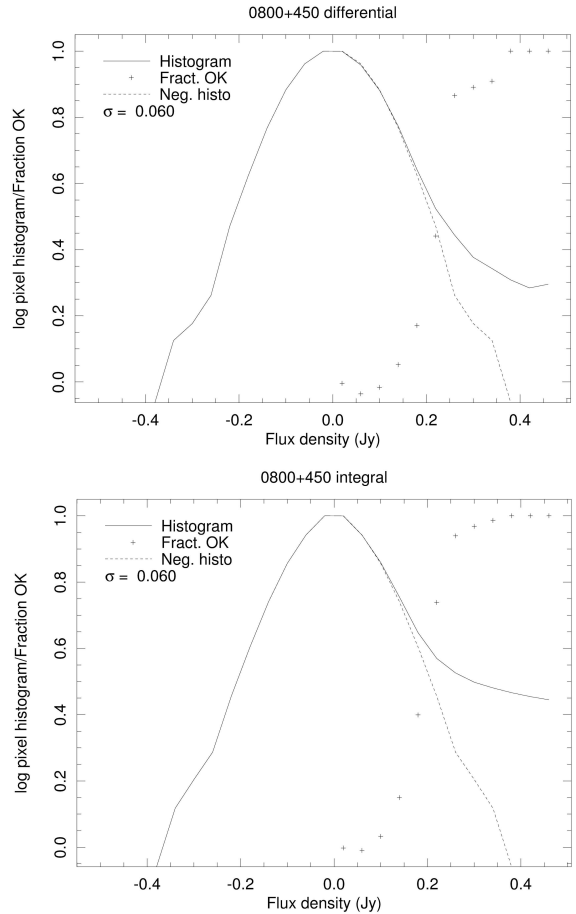}
\caption{Differential ({\it top}) and integrated ({\it bottom})
  histograms of the pixel values for a VLSSr square.  The solid line
  shows the histogram.  The dotted line shows the negative-pixel value
  histogram projected onto the positive bins.  The plus-signs show the
  fraction of pixels which represent real sources at a given flux
  value, assuming the distribution is symmetric.  \label{fig:pixhist} }
\end{figure}

\section{Improvements to Analysis Techniques}
\subsection{False Detection Rate Cataloging}

Wide-field astronomical images, particularly those intended as sky
surveys, are typically decomposed into a catalog of objects.  However,
pixel values in astronomical images always contain a randomly
distributed component, which is unrelated to anything on the celestial
sphere.  Some criterion must be adopted to distinguish between
features in the image which are a result of this ``noise'', and thus
unlikely to be real, and sources which do correspond to real objects.
The chosen criterion will always involve a trade-off between the
possibility of missing real sources and the contamination of the final
catalog by false sources.

For cases where the noise has a Gaussian distribution, tests for the
statistical probability of any feature being due to the noise
distribution are well established.  A common, and simple choice, is to
make a cutoff at some multiple of the RMS, or $\sigma$ of the
distribution.  More sophisticated algorithms for images with Gaussian
noise have also been developed ~\citep{Hopkins2002, Friedenberg2009}.
However, low frequency radio images, such as those that form the VLSS,
do not have noise with a Gaussian distribution.  The poorly known
primary antenna pattern and difficulties with modeling the effects of
ionospheric fluctuations result in a non-trivial fraction of the
celestial power being scattered into fake features.  As a result the
Gaussian statistics underestimate the number of false detections.

An alternative approach is to estimate the false detection probability
directly from the image statistics. If we create a pixel distribution
for an image, the negative tail should represent some combination of
thermal noise, calibration and imaging artifacts.  The positive tail
represents those plus real sources.  If we assume that the noise
should be symmetric, we can estimate the true positive noise from
the negative half.  The ratio of the excess positive values in a
positive flux bin to the negative values in the corresponding negative
flux bin equals the fraction of positive values likely to be real
sources at that flux value.

$$FDR_x\ =\ 1\ -\ {n_+-n_-\over{n_-}}$$ 

where $FDR_x$ is the false detection rate at flux density level $x$,
$n_+$ is the number of pixels in the positive $x$ bin and $n_-$ is the
number of pixels in the negative $x$ bin.  In the top panel of Figure
3, a sample histogram of pixel values from one of the VLSSr squares is
plotted.

In order to use this method successfully, good statistics well out
into the wings of the distribution are needed; this translates to
sampling large numbers of pixels.  If the character of the noise
changes across the map it may be preferred to create statistics over
more limited areas.  For a survey such as the VLSS, where each map is
a mosaic of overlapping pointings, the statistical properties of the
noise can be extremely variable.

To make the statistics more robust for smaller numbers of pixels, an
integrated pixel distribution can be used; thus each flux bin includes
counts of all pixels at that flux or any flux further from zero.  The
difference in the two distributions can be seen in Figure 3.  The
calculated false detection rate can then be stated as the probability
that a pixel at a given flux or greater is real.

This method allows the person generating the catalog to choose the
target false detection rate when making the catalog.  When using the
resulting catalog, the number of false sources is theoretically known.
False detection rate (FDR) cataloging has been implemented in the Obit
task FndSou, and can be run on a map subsection of arbitrary size.
More details can be found in ~\citet{FDR}.

In order to test the effectiveness of FDR cataloging on the VLSSr we
compared the cataloged sources to the much more sensitive NVSS; we
would expect only a small fraction of real sources to appear in the
VLSS and not the NVSS.  Unfortunately, the noise distribution in the
VLSSr is not symmetric; there is an excess of positive sidelobes in
many areas on the sky.  The false detection rate method produced a
catalog with $10\%$ more sources than traditional $5\sigma$ cataloging
tests; however nearly half of those additional sources were fake.
When targeting a $1\%$ false detection rate, we achieved closer to a
$9\%$ rate, considerably higher than the $5\%$ false detection rate we
found using traditional Gaussian ($5\sigma$) thresholding.  We thus consider
the $5\sigma$ catalog to be the VLSSr ``final'' catalog.

\begin{figure*}[t]
\noindent\includegraphics[width=39pc]{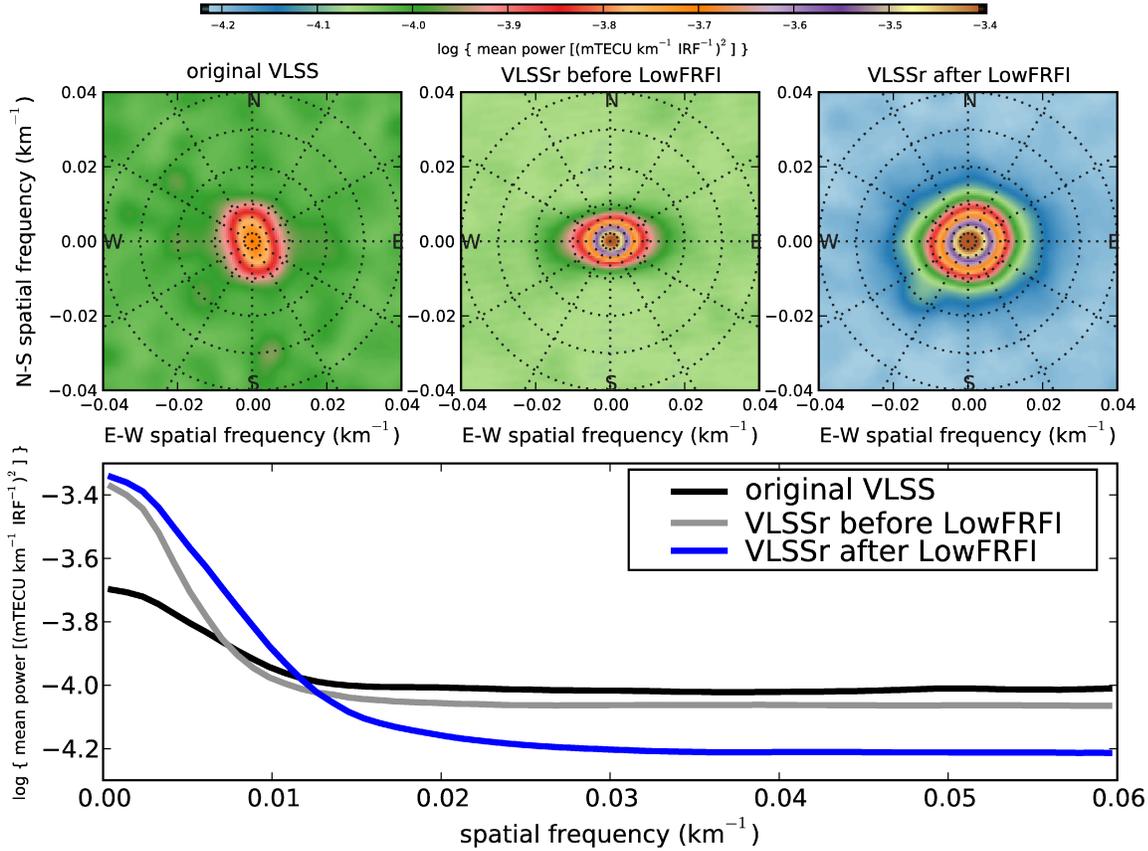}
\caption{{\it Top Panel:} Average power spectra of the fluctuations in
  total electron count (TEC) in two dimensions for the VLSS ({\it left}),
  the VLSSr before RFI-mitigation ({\it middle}) and the final VLSSr
  ({\it right}).  Data are derived from the calibrator source offsets
  measured for the field-based calibration.  {\it Bottom Panel:} A
  radial representation of the power spectrum for the same three
  data sets.  The VLSSr shows an increase in power compared to the VLSS
  at the smallest spatial frequencies (largest wavelengths).  It also
  has a much lower noise floor allowing the fluctuations to be probed
  to ionospheric spatial scales that are a factor of 2 larger
  (corresponding to smaller wavelengths) compared to the
  VLSS. \label{fig:ionfig} }
\end{figure*}

\subsection{Ionospheric Fluctuation Analysis \label{sec:ion} }

The position offset data used within the field-based ionospheric
correction of the VLSSr contains a wealth of information about the
ionospheric fluctuations present during the observations.
~\citet{CohenION} produced a statistical analysis of these
fluctuations using the offset data from the original VLSS reduction.
Using what are essentially TEC gradient structure functions, they
demonstrated that the median behavior of the ionosphere was roughly
turbulent, with substantially more activity during the day than at
night.  The analysis was hampered both by the original VLSS reduction
software and by the ability of the structure functions to characterize
individual disturbances.

New software has been recently developed which performs a
Fourier-based analysis of the position offset data for all calibrator
sources during each observation to produce a three-dimensional (one
temporal and two spatial) power spectrum ``cube'' of TEC gradient
fluctuations.  These cubes provide a statistical description of the
ionospheric environment and allow the identification and
characterization of transient phenomena.  The technique is described
in detail in \citet{ION1}.

The new software has been applied to the position shifts found in the
ionospheric correction step of both the VLSS and the VLSSr; in the
latter case information is available both before and after the
RFI-mitigation step and both were analyzed.  We present a brief
overview of the results here; detailed results of this analysis will
be presented in a companion paper, \citet{ION2}.  The top panel in
Figure 4 shows the mean two-dimensional power spectrum of fluctuations
in the total electron count (TEC) gradient of the ionosphere, while
the bottom shows the azimuthally averaged spectra for each of the
three data sets.  The power spectra are smoothed by the time sampling
of 1 to 2 minutes.  We have also assumed a single gradient across the
array, which smooths the measurements with an 11-km wide kernel (the
diameter of the VLA B-configuration).  This corresponds to a
sinc$^{2}$-shaped taper of the power spectra which goes to zero at a
spatial frequency of 1/11 km$^{-1}$ or 0.091 km$^{-1}$.  Please see
\citet{ION2} for more details.

There is a dramatic increase in power on the largest scales (smallest
spatial frequencies) between the original reduction, which used the
NVSS input catalog, and the current reduction, which uses the VLSS
itself for input.  As described in Section ~\ref{sec:ioncalr}, this is
a reflection of our ability to include sources at larger positional
shifts in our ionospheric models, which in turn is a result of having
a proper sky model at the data frequency.  The larger position shifts
correspond to larger-amplitude fluctuations. Note that this also
reflects how the improved reduction technique works under more adverse
circumstances than previously.  Higher average power at low spatial
frequencies means on the average a more disturbed ionosphere in the
data included.

The accuracy of the the TEC gradient measurement is roughly
proportional to the uncertainty of the position offsets which improves
with lower noise.  So the radial spectrum of the VLSS flattens out due
to noise at a spatial frequency of about 0.015 km$^{-1}$,
corresponding to wavelengths less than $\sim$70 km.  While the pre-RFI
mitigation VLSSr is similarly limited due to noise, there is a
dramatic improvement after the RFI mitigation step, which lowered the
noise.  The post-mitigation spectrum flattens out around spatial
frequencies of 0.03 km$^{-1}$ (wavelengths of $\sim 35$km), increasing
the range of ionospheric structure scales that can be probed by a
factor of two.

\section{Conclusions} 

We have recently reprocessed the VLSS data to create a revised version
of the survey, which we call the VLSSr.  This new reduction took
advantage of improvements to the data reduction process, including an
improved peeling algorithm, smart-window cleaning to reduce clean
bias, higher order Zernike models to correct ionospheric effects, and
RFI modeling techniques.  We also investigated a new source cataloging
criterion and were able to make an improved and expanded ionospheric
analysis based on the ionospheric Zernike model calculations.  All of
the improved algorithms except the ionospheric analysis software are
available in the Obit data reduction package.

Although the VLSSr provides a substantial improvement over the VLSS
for much of the sky, we were unable to image data from the previously
unpublished low declination areas centered near 18 hrs in Right
Ascension.  These regions were observed twice because the first data
were corrupted by extreme ionospheric weather which could not be
adequately modeled by the Zernike polynomials.  Unfortunately the
re-observations were affected by instrumental problems during the
recent VLA upgrade, and also cannot be imaged reliably.

Roughly 5$\%$ of the remaining fields did not improve with the new
reduction techniques.  Often these fields exhibit signs of distorted
sources just outside the field of view, and/or poor primary
calibration.  In the original VLSS we self-calibrated many fields to
mitigate these types of issues before applying field-based
corrections.  Because the number of affected fields was so small we
chose not to re-introduce the self-calibration step for the VLSSr.
The images and source catalogs for these fields are included in the
final survey products.

The improved reduction techniques allowed us to image six previously
unpublished fields (1$\%$), most near extremely strong sources such as
Cassiopeia A.  The final catalog includes approximately 95,000 source
components, of which 74,000 are unresolved.  Sources were fitted with
Gaussians which could have maximum sizes of 120''; larger sources were
fitted with multiple Gaussians.  In the published VLSS catalog
multiple-component sources were summed to create one entry; we have
chosen to leave the individual component entries uncombined for the
VLSSr.  

Comparing the VLSSr to the VLSS, the clean bias was reduced by over
50$\%$, the largest angular scale imaged was aproximately doubled, and
the number of cataloged sources increased by 35$\%$.  We decreased the
restoring beam size from $80''$ to $75''$, but average errors on the
source positions increased slightly (from $\sim 3''$ to $\sim 3.4''$
in RA and Dec).  The new reduction doubles the range of spatial scales
over which we are able to measure the power spectrum of ionospheric
fluctuations.

%%% End of body of article:

%%%%%%%%%%%%%%%%%%%%%%%%%%%%%%%%
%% Optional Appendix goes here
%
% \appendix resets counters and redefines section heads
% but doesn't print anything
% After typing  \appendix
%
% \section{Here Is Appendix Title}
% will show
% Appendix A: Here Is Appendix Title
%
%%%%%%%%%%%%%%%%%%%%%%%%%%%%%%%%%%%%%%%%%%%%%%%%%%%%%%%%%%%%%%%%
%
% Optional Glossary or Notation section, goes here
%
%%%%%%%%%%%%%%
% Glossary is only allowed in Reviews of Geophysics
% \section*{Glossary}
% \paragraph{Term}
% Term Definition here
%
%%%%%%%%%%%%%%
% Notation -- End each entry with a period.
% \begin{notation}
% Term & definition.\\
% Second term & second definition.\\
% \end{notation}
%%%%%%%%%%%%%%%%%%%%%%%%%%%%%%%%%%%%%%%%%%%%%%%%%%%%%%%%%%%%%%%%
%
%  ACKNOWLEDGMENTS

\begin{acknowledgments}
We would like to thank the referees for helpful questions, comments and suggested revisions.  Basic research in radio astronomy at the Naval Research Laboratory is supported by 6.1 base funds. The National Radio Astronomy Observatory
is a facility of the National Science Foundation operated under
cooperative agreement by Associated Universities, Inc.

\end{acknowledgments}

\end{article}

%% Enter Figures and Tables here:
%\vfill\eject

% When submitting articles through the GEMS system:
% COMMENT OUT ANY COMMANDS THAT INCLUDE GRAPHICS.
%
% FOR FIGURES, DO NOT USE \psfrag or \subfigure commands.
%
% Figure captions go below the figure.
% Table titles go above tables; all other caption information
%  should be placed in footnotes below the table.

% DRAFT figure/table, including eps graphics
%
% \begin{figure}
% \noindent\includegraphics[width=20pc]{samplefigure.eps}
% \caption{Caption text here}
% \end{figure}
% \end{document}
%
% \begin{table}
% \caption{}
% \end{table}
%

% ---------------
% TWO-COLUMN figure/table
%
% \begin{figure*}
% \noindent\includegraphics[width=39pc]{samplefigure.eps}
% \caption{Caption text here}
% \end{figure*}
%
% \begin{table*}
% \caption{Caption text here}
% \end{table*}
%
% ---------------
% EXAMPLE TABLE
%
%\begin{table}
%\caption{Time of the Transition Between Phase 1 and Phase 2\tablenotemark{a}}
%\centering
%\begin{tabular}{l c}
%\hline
% Run  & Time (min)  \\
%\hline
%  $l1$  & 260   \\
%  $l2$  & 300   \\
%  $l3$  & 340   \\
%  $h1$  & 270   \\
%  $h2$  & 250   \\
%  $h3$  & 380   \\
%  $r1$  & 370   \\
%  $r2$  & 390   \\
%\hline
%\end{tabular}
%\tablenotetext{a}{Footnote text here.}
%\end{table}

% See below for how to make landscape/sideways figures or tables.

\end{document}